\documentclass[a4paper,aps,prd,superscriptaddress,nofootinbib,notitlepage,twocolumn,floatfix]{revtex4-1}
\usepackage{graphicx}
\usepackage{amssymb,amsmath,amsfonts,nccmath}
\usepackage[normalem]{ulem}

\usepackage{color}
\usepackage{dsfont}
\usepackage{mathrsfs}

\newcommand \be{\begin{equation}}
\newcommand \ee{\end{equation}}
\newcommand \bea{\begin{eqnarray}}
\newcommand \eea{\end{eqnarray}}

\newcommand{\MeV}{{\rm MeV}}
\def\Li{\textrm{Li}}

\begin{document}

\author{Attila P\'asztor}
\email{apasztor@bodri.elte.hu}
\affiliation{ELTE E{\"o}tv{\"o}s Lor\'and University, Institute for Theoretical Physics, P\'azm\'any P.~s.~1/A, H-1117, Budapest, Hungary.}

\author{Zsolt Sz{\'e}p}
\email{szepzs@achilles.elte.hu}
\affiliation{MTA-ELTE Theoretical Physics Research Group, P\'azm\'any P.~s.~1/A, H-1117 Budapest, Hungary.}

\author{Gergely Mark{\'o}}
\email{gmarko@physik.uni-bielefeld.de}
\affiliation{Fakult{\"a}t f\"ur Physik, Universit{\"a}t Bielefeld, D-33615 Bielefeld, Germany.}

\title{Apparent convergence of Pad{\'e} approximants for the \\ crossover line in finite density QCD}

\begin{abstract}
We propose a novel Bayesian method to analytically continue observables to real baryochemical potential $\mu_B$ in finite density QCD. Taylor coefficients at $\mu_B=0$ and data at imaginary chemical potential $\mu_B^I$ are treated on equal footing. We consider two different constructions for the Pad{\'e} approximants, the classical multipoint Pad{\'e} approximation and a mixed approximation that is a slight generalization of a recent idea in Pad{\'e} approximation theory. Approximants with spurious poles are excluded from the analysis. As an application, we perform a joint analysis of the available continuum extrapolated lattice data for both pseudocritical temperature $T_c$ at $\mu_B^I$ from the Wuppertal-Budapest Collaboration and Taylor coefficients $\kappa_2$ and $\kappa_4$ from the HotQCD Collaboration. An apparent convergence of $[p/p]$ and $[p/p+1]$ sequences of rational functions is observed with increasing $p.$ We present our extrapolation up to $\mu_B\approx 600$~\MeV.
\end{abstract}

\maketitle

\section{Introduction}

Despite considerable effort invested so far, the phase diagram of QCD in the temperature($T$)-baryon chemical potential($\mu_B$) plane still awaits determination from first principles. At the moment, the only solid information available is the curvature of the crossover temperature~\cite{Bonati:2018nut, Bazavov:2018mes, Borsanyi:2020fev}, together with some upper bound on the absolute value of the next Taylor coefficient of order $\mathcal{O}(\mu_B^4)$~\cite{Bazavov:2018mes, Borsanyi:2020fev}. These results come either from the evaluation of Taylor coefficients with lattice simulations performed at $\mu_B=0$ or via simulations performed at imaginary $\mu_B$, where the sign problem is absent, with the Taylor coefficients obtained from a subsequent fit.

Whether the input data are the Taylor coefficients or the values of a function at several values of the imaginary chemical potential, fact is that the numerical analytic continuation needed to extrapolate the crossover to real $\mu_B$ is a mathematically ill-posed problem \cite{Tikhonov,Stef}. This means that although the analytic continuation of a function sampled inside some domain $D$ is uniquely determined by the approximant used, the extension of a function differing on $D$ by no matter how small an amount can lead to arbitrarily different values at points outside $D$. That is to say: analytic continuation is unique, but is not a continuous function of the data. For such ill-posed problems, the only way to achieve convergence in the results is to use some kind of regularization. This makes sure that the noise in the data is not overemphasized by the analytic continuation. As the noise is reduced, the regularizing term is made weaker. This leads to a kind of double limit when the regularization and the noise are taken to zero together. The simplest kind of regularization for analytic continuation is the use of some ansatz, which is assumed to describe the physics both in the range where data is available, and in the range where one tries to extrapolate. The conservative view is to use for analytic continuation few-parameter approximants, which all fit the data well, and perform the continuation only in a range where they do not deviate much from each other, assessing the systematic error of the continuation from this deviation. Here we pursue a more adventurous approach, by considering a sequence of approximants of increasing functional complexity, and trying to observe whether they converge or not.

In the absence of physically motivated ansatz, a good guess is to study the $[p/p]$ (diagonal) and $[p/p+1]$ (subdiagonal) Pad{\'e} sequences, as these are only slightly more complicated to work with than polynomials, but have far superior convergence properties. Ordinary Pad{\'e} approximants (i.e. rational functions constructed using approximation-through-order conditions to match the Taylor expansion of a function at a given point) are known to converge uniformly on the entire cut plane for functions of Stieltjes type \cite{Bender} (which have a cut on the negative real axis). For this class of functions the subdiagonal sequence of multipoint (or $N-$point) Pad{\'e} approximants \cite{Baker}, also know as the Schlessinger point method in the context of scattering theory \cite{Schlessinger}, is also convergent (see \cite{Gelfgren} and references therein). For a meromorphic function, on the other hand, Pad{\'e} approximants are known to converge in measure~\cite{Nuttall,ZinnJustin:1971a}, i.e. almost everywhere on the complex plane, in stark contrast to polynomial approximations, which stop converging at the first pole of such a function.

While the convergence properties of Pad{\'e} approximants in exact arithmetic are often very good, even in cases where the mathematical reason for the convergence is not fully understood yet, these approximations tend to be very fragile in the presence of noise. This often manifests itself in spurious poles, whose residue goes to zero as the noise level is decreased, as well as spurious zero-pole pairs (called Froissart doublets~\cite{Froissart}). The distance between the zero and the pole goes to zero as the noise decreases, eventually leading to the annihilation disappearance of the doublet. There is a large body of mathematical literature devoted to the removal of these spurious poles. Procedures which do so typically involve some further regularization, like in Ref.~\cite{Gonnet}, where this is based on singular value decomposition, or monitoring the existence of Froissart doublets for later removal, like in Ref.~\cite{Beckerman}. In cases where the noise level on the data cannot be arbitrarily decreased~\footnote{E.g. if the noise is only coming from machine precision in floating point arithmetic, the Froissart doublets can often be removed by simply using multiple precision arithmetic for the ``naive'' algorithm.}, the exclusion of spurious poles is mandatory if one wants to go to higher order approximants in the analysis.

When dealing with numerical analytic continuation, we need to select the approximant from a class of possible functions (i.e. a model) and a method to take into account the data (i.e. a fitting method). For the former we use  two types of rational approximants, the classical multipoint Pad{\'e} approximants recently used for analytic continuation in Refs.~\cite{Pilaftsis:2013xna, Marko:2017yvl, Tripolt:2018xeo} and a slight generalization of the Pad{\'e}-type approximant introduced and studied recently in \cite{Brezinski}. The parameters of the multipoint Pad{\'e} approximant are determined solely in terms of the interpolating points and information on Taylor coefficients, if it exists, can be taken into account in the second, data fitting step. In contrast, the Pad{\'e}-type approximant allows for a joint use of interpolating points and Taylor  coefficients in determining the parameters of the approximant.  Although the focus in \cite{Brezinski} was on the diagonal sequence $[p/p]$ of Pad{\'e}-type approximants, the method can be easily generalized to construct the subdiagonal sequence as well. For the data fitting step we use a Bayesian analysis. The likelihood function ensures that approximants are close to both the data on the Taylor coefficients at $\mu_B=0$ and the data at purely imaginary $\mu_B$, while a Bayesian prior makes sure that spurious poles are excluded from the extrapolation. Considering two different types of Pad{\'e} approximants is a nontrivial consistency check, mainly because the exact form of the prior distribution will be different for the two cases, as the number of interpolation point where the function values will be restricted is different.

We note that while Bayesian methods---especially variations of the maximum entropy method with different entropy functionals---for the analytical continuation to real time are quite commonly used in lattice QCD~\cite{Asakawa:2000tr, Jakovac:2006sf, Aarts:2007wj, Meyer:2011gj, Rothkopf:2011ef, Burnier:2013nla, Borsanyi:2014vka, Rothkopf:2019dzu}, as far as we are aware, such methods have not been applied to the analytic continuation problem in $\mu_B$ so far. The only related example we are aware of is Ref.~\cite{Borsanyi:2018grb}, where a Bayesian method is used to extract high order derivatives of the pressure around $\mu_B=0$ from data at imaginary $\mu_B$. One must note however, that the mathematical  problem in that paper is that of numerical differentiation, which is distinct from the analytic continuation problem discussed here. This paper, therefore, is the first attempt of using this class of mathematical techniques to the analytic continuation problem in finite density QCD.

The above mentioned fragility of the Pad{\'e} approximation method when applied to noisy data is the main reason that most of the previous applications to finite density QCD employ low order approximants. Pad{\'e} approximants were used in this context to analytically continue to real values of $\mu_B$ the pseudocritical temperature values obtained at imaginary chemical potential for various number of flavors and colors \cite{Lombardo:2005ks, Cea:2009ba, Cea:2012ev, Bellwied:2015rza}. The convergence of a Pad{\'e} sequence was seemingly not in the focus of these investigations, with the exception of \cite{Lombardo:2005ks}. A related problem in finite density QCD, where Pad{\'e} approximants have also been considered, is the calculation of the equation of state at finite chemical potential. An early work that uses a high order Taylor expansion in an effective model is Ref.~\cite{Karsch:2010hm}. Two recent examples in lattice QCD are Refs.~\cite{Datta:2016ukp,Gunther:2016vcp}. The low order Pad{\'e} approximants used in the above studies are not yet expected to take advantage of the superior convergence properties of the Pad{\'e} series. This is in sharp contrast to the case in statistical physics, where in Ising-like models the Taylor coefficients are known exactly to high orders \cite{ising1,ising2,ising3}. However, even a low order Pad{\'e} approximant represents a resummation of the Taylor series, which is exploited when applied outside the radius of convergence of the Taylor series. The main advantage of the Bayesian approach presented here is the ability to go to considerably higher orders, at the cost of what we believe are physically reasonable extra assumptions.

The paper is organized as follows. In Sec.~\ref{sec:method} we introduce the mathematic tools used for our analysis. First we treat the novel Pad{\'e}-type approximants in the absence of noise. Since the traditional multipoint Pad{\'e} approximants are quite well known, they are relegated to Appendix~\ref{app:mP}. Next, we discuss the Bayesian analysis in the presence of noise in a general manner that includes both the multipoint Pad{\'e} and the mixed Pad{\'e} approximant case. In Sec.~\ref{sec:Tc-cont} we turn to physical applications. We first demonstrate the effectiveness of Pad{\'e} approximants in a chiral effective model. Finally, we perform a joint analysis of the continuum extrapolated lattice data on the Taylor coefficients at $\mu_B=0$ and the crossover line at imaginary $\mu_B$. Appendix B summarizes the formulas relating our notational conventions on the Taylor coefficients to those found elsewhere in the literature.

\section{Numerical method for analytic continuation \label{sec:method}}

\subsection{Pad{\'e}-type rational approximants in the absence of noise}

Using the notation of \cite{Brezinski} the mathematical formulation of analytic continuation is as follows. Assuming the existence of a continuous real function $f:\mathbb{R}\to\mathbb{R},$ we would like to know its value for $t>0$ given that:
\begin{enumerate}
\item
  at a number of \emph{interpolating points} $\tau_i<0$, $i=1,\dots,l,$ the values $f_i:=f(\tau_i)$ are known,
\item
  a number of coefficients $c_i$, $i=0,\dots,k$ in the Taylor expansion
  \be
  \label{Eq:Taylor_exp}
  f(t)=c_0+c_1 t+\dots +c_k t^k
  \ee
  around $t=0$ are known.
\end{enumerate}

A widely used method to tackle the problem is to fit a rational fraction\footnote{One can choose $b_0=1$ without loss of generality, having $n+m+1$ independent coefficients.}
\be
\label{Eq:rational_func}
   [n/m] \equiv R_m^n(t) \equiv \frac{N_n(t)}{D_m(t)} := \frac{\displaystyle \sum_{i=0}^n a_i t^i}{\displaystyle \sum_{i=0}^m b_i t^i,}
\ee
to the set of function values and/or available derivatives. When only derivatives at $t=0$ are used, one obtains the \emph{ordinary Pad{\'e} approximant} in which the coefficients of both the denominator and the numerator are fully determined by the Taylor coefficients by imposing the approximation-through-order conditions\footnote{One equates the expansion
of $f(t)$ with $R_m^n(t)$, cross multiply and then equates the coefficients
of $t$ on both sides of the equation.} $R_m^n(t)=f(t)+{\cal O}(t^{m+n+1}).$ This condition implies that the Taylor expansion of the Pad{\'e} approximant around $t=0$ agrees with the Taylor expansion of the function up to and including the order of the highest Taylor coefficient known.

When only the set of function values at the interpolating points are used, one obtains the so-called \emph{multipoint Pad{\'e} approximant}, which is particularly useful in numerics in its continuous fraction formulation because the coefficients can be determined easily from recursion relations. For odd number of points $N=2k+1,k\ge 0,$ one obtains the approximant $[k/k]$, while for even number of points $N=2k, k\ge 1$, one obtains the approximant $[k+1/k].$ More details can be found in Appendix~\ref{app:mP}, where the construction of the multipoint approximant $C_N$ is summarized.

As mentioned in Ref.~\cite{Baker-Gammel} (see p.~16), an obvious modification of the multipoint Pad{\'e} approximation can be given if any number of successive derivatives exists at the points where the value of the function is known. Recently such a modification, called \emph{Pad{\'e}-type rational approximant} with $n=m=k$ was constructed in \cite{Brezinski}, with $k$ being the degree up to (and including) which the expansion of the approximant matches the Taylor expansion of the function. The denominator of this  $[k/k]$ approximant is fixed by function values at arbitrarily chosen interpolating points and the coefficients of the numerator are obtained by imposing the approximation-through-order conditions.

It is easy to generalize the construction used in \cite{Brezinski} to obtain Pad{\'e}-type approximants for which $l\ne k.$ With $k+1$ coefficients of the Taylor expansion (including the value of the function at zero) one can construct many Pad{\'e}-type approximants of this type, one
just has to satisfy the relation $n+m=k+l$. In this case $k+1$ coefficients of the numerator $N_n(t)$ are determined from the approximation-through-order conditions, meaning that strictly speaking $R_m^n$ satisfies by construction $n\ge k,$ and the remaining $n+m-k=l$ coefficients are fixed by function values at $l$ number of interpolating points via $f_i D_m(\tau_i)=N_n(\tau_i), i=1,\dots,l.$

In what follows we shall use Pad{\'e}-type approximants of the form $R_p^p$ and $R_{p+1}^p,$ with $p\ge 1$, satisfying $2p=k+l$ and $2p+1=k+l,$ respectively. To construct for example $R_4^3(t)$ using $c_0,c_1,$ and $c_2$, one equates \eqref{Eq:Taylor_exp} with \eqref{Eq:rational_func} and after cross multiplication one matches the coefficients
of $t^0,t^1,$ and $t^2$. This gives $a_0=c_0,$ $a_1=c_1+b_1c_0,$ and $a_2=c_2+b_1c_1+b_2c_0,$ which are common for all approximants with $n\ge2.$ Using these expressions for $a_0,a_1$, and $a_2$ in $N_3(t)$, one sees that the five conditions $f_i D_4(\tau_i)=N_3(\tau_i),i=1,\dots,5$ represents a system of linear equations for the five unknown $a_3,b_1,b_2,b_3,b_4,$ which can be easily solved numerically with some standard linear algebra algorithm. As for the $R_2^1(t)$ approximant, this is constructed very similarly to the original Pad{\'e} approximant, only the condition on the third derivative (unknown in our case) is replaced by $R_2^1(\tau_1)=f_1,$ where $\tau_1$ is an interpolating point.

\subsection{Bayesian approach \label{ss:bayes}}
The Bayesian approach \cite{murphy} that considers the data sample fixed and the model parameters as random variables gives a perspective on the curve fitting
problem which is particularly suited for a meta-analysis of data with noise.

We do not include Pad{\'e} approximants of different order in one large meta-analysis, rather we perform a separate Bayesian analysis of the
different order approximants, in order to study their convergence properties as the order of the approximation is increased.
For an $[n/m]$ Pad{\'e} approximant, the model parameters are the coefficients $\vec{a}=(a_0,a_1,\dots,a_n)$ and $\vec{b}=(b_1,b_2,\dots,b_m)$,
with a total of $n+m+1$ coefficients to be determined.
The posterior probability can be written as:
\begin{equation}
  \mathcal{P}(\vec{a},\vec{b}| \textrm{data})
  = \frac{1}{Z}
  \mathcal{P}(\textrm{data} | \vec{a},\vec{b})
  \mathcal{P}_{\textrm{prior}}(\vec{a},\vec{b}) \rm{,}
\end{equation}
where assuming Gaussian errors around the correct model parameters, the likelihood is given by:
\begin{equation}
  \begin{aligned}
  \mathcal{P}(\textrm{data} | \vec{a},\vec{b}) &= \exp \left( -\frac{1}{2} \chi^2 \right) \rm{,} \\
      \chi^2 &= \chi^2_\textrm{Taylor} + \chi^2_{\rm{Im} \mu_B} \rm{,}\\
      \chi^2_\textrm{Taylor} &= \sum_{i=1}^{T} \frac{\left(
      c_i - \frac{\partial^{i} R^n_m(\mu_B;\vec{a},\vec{b}) }
                  {\partial \left(\mu^2_B\right)^{i}} \Big|_{\mu_B=0}
      \right)^2}{\sigma_{c_i}^2} \rm{,}\\
      \chi^2_{\rm{Im} \mu_B} &= \sum_{j=1}^{L} \frac{\left(
        f_j - R^n_m (i\mu_{B,j}^I;\vec{a},\vec{b})
      \right)^2}{\sigma_{f_j}^2} \rm{,}\\
  \end{aligned}
  \label{Eq:PP}
\end{equation}
with $T$ being the number of derivatives known at $\mu_B=0$ and $L$
being the number of function values known for $\mu_B^2<0$.  $Z$ is a
normalization constant. The Taylor
coefficients at $\mu_B=0$ are clearly correlated, but their correlation
matrix was not given in Ref.~\cite{Bazavov:2018mes} so we ignore the
correlations. If the correlations between the Taylor coefficients are
known, including them in our method is completely straightforward.
The data at different values of imaginary $\mu_B$ come from
different Monte Carlo runs, and are thus uncorrelated.

The variables that the Bayesian analysis code uses for the construction of the
Pad{\'e} approximants are not the coefficients of the polynomial themselves. For the multipoint
Pad{\'e} approximants we use a number of interpolated values at fixed node points in $\hat\mu_B^2:=\mu_B^2/T^2$.
For the case of Pad{\'e}-type approximants we use a smaller number of interpolated values at node
points and a number of derivatives at $\mu_B=0$. These are of course in a one-to-one correspondence with
the polynomial coefficients, once the restriction $b_0=1$ has been made in Eq.~\eqref{Eq:rational_func}.
Details of the implementation will be discussed in Sec.~\ref{sec:Tc-lattice}.

An important part of our procedure is that we do not work with the space of all Pad{\'e}
approximants of order $[n/m]$, rather, the allowed approximants are  restricted by the prior, which
always contains a factor that excludes spurious poles both in the interpolated and the extrapolated range.
Due to this factor of the prior, the method is only applicable when no physical poles are
expected in the aforementioned ranges.

The prior also contains a further factor---the exact form of
which for the two different Pad{\'e} approximants will be discussed in Sec.~\ref{sec:Tc-lattice}---which
prevents extra oscillations of the interpolants in the $\mu_B^2<0$ range, which are
not warranted by the data. This is enforced by using a prior distribution of the
interpolated values at the node points at fixed $\mu_B^2/T^2$ range. We have checked that our results
are not sensitive to the choice of the node points. This is also expected on mathematical grounds, since
unlike polynomial interpolants, rational interpolants are not extremely sensitive to the choice of the
node points used for the interpolation
~\cite{ATAP}.

Putting all the above information together, the prior can be given as an implicit condition on the model parameters $\vec{a}$ and $\vec{b}$ in the following form
\begin{align}
&\mathcal{P}_{\textrm{prior}}(\vec{a},\vec{b})\nonumber\\
&\propto
\begin{cases}
\begin{array}{ll}
    \displaystyle \prod_\textrm{i} F(|R_m^n(i\mu_{B,\textrm{i}}^I;\vec{a},\vec{b})-\bar T_c(\hat\mu^2_{B,\textrm{i}})|,w_\textrm{i}), & \nexists\ \textnormal{pole} \in \mathcal{I},\\
    0, & \exists\ \textnormal{pole} \in \mathcal{I},
\end{array}
\end{cases}
  \label{Eq:Pprior}
\end{align}
where $F(x,w)$ is either $\exp{(-x^2/(2 w^2))}$ or $\theta(w-x)$ (Heaviside step function), corresponding, respectively, to \emph{Method 1} and \emph{Method 2} used in Sec.~\ref{ss:Bayes-implement} and $\mathcal{I}$ represents a range of $\hat\mu_B^2$ for which the absence of poles of the Pad{\'e} approximants is required (we use $\mathcal{I}=[-\pi^2,60\pi^2]$). The index ``$\textrm{i}$'' goes over the interpolating (node) points, which are different from the data points $j=1,\dots,L$ used in \eqref{Eq:PP}. In \emph{Method 1} the temperature values $R_m^n(i\mu_{B,\textrm{i}}^I;\vec{a},\vec{b})\equiv T_c^\textrm{i}$ at the node points are generated with a normal distribution whose standard deviation $w_i$ is chosen to be substantially larger than the error $\sigma_{f_j}$ of the lattice data, in which case the result of the analytic continuation is not sensitive to the actual value of $w_i.$ In \emph{Method 2} the temperature values at the node points are generated using importance sampling and the $\theta$-function is only needed for a technical reason, as explained at the end of Sec.~\ref{ss:Bayes-implement}. The node points, as well as $w_\textrm{i}\equiv w_{T_c^\textrm{i}}$ and $\bar T_c(\hat\mu^2_{B,\textrm{i}})$, are given in Fig.~\ref{Fig:prior}. $\bar T_c(\hat\mu^2_{B,\textrm{i}})$ is obtained by interpolating the mean value of the lattice data points available at imaginary $\mu_B$.

Our numerical results will be based on the posterior distribution. For a fixed value of $\mu_B/T$, we study the posterior distribution of the crossover temperature $T_c=R_m^n(\hat\mu_B^2)$ and chemical potential $\mu_B=\sqrt{\hat\mu_B^2}T_c$. The center point will in both cases be the median, while the asymmetric error bars represent the central $68\%$ of the posterior distribution of both quantities. We will call these \emph{percentile based errors}. We shall see that the asymmetry of the posterior distribution increases as $\hat\mu_B^2$ increases in the extrapolation range (real values of $\mu_B$) and is the largest for the [2/2] Pad{\'e} approximant. In practice, the integration over the prior distribution is carried out with simple Monte Carlo algorithms. The statistics needed is such that the posterior distribution of the studied observables does not change anymore, which we explicitly checked to be the case in our analysis.

\section{Analytic continuation of $T_c(\mu_B^I)$ \label{sec:Tc-cont}}

\subsection{Convergence of Pad{\'e} approximants in a chiral effective model \label{ss:model}}

Before applying the method described in Sec.~\ref{sec:method} to the actual QCD data, we study the analytic continuation within the chiral limit of the two flavor ($N_f=2$) constituent quark-meson (CQM) model.
We show that in this model both the diagonal and the subdiagonal sequences constructed from $T_c(\mu_B^I)$ exhibit apparent convergence to the exact $T_c(\mu_B)$ curve. We also show that the Pad{\'e} approximant
knows nothing about the location of the tricritical point (TCP), as this information is not encoded in $T_c(\mu_B).$ Finally, we investigate the effect of the error on the analytical continuation.

\subsubsection{Convergence in the absence of noise}

In Ref.~\cite{Jakovac:2003ar}, using leading order large-$N$ techniques resulting in an ideal gas approximation for the constituent quarks, the coefficients of the Landau-Ginzburg type effective potential $V_\textrm{eff}=\frac{m^2_\textrm{eff}}{2}\Phi^2+\frac{\lambda_\textrm{eff}}{4}\Phi^4+\dots$ for the chiral order parameter $\Phi$ were determined in the chiral limit to be\footnote{These expressions corresponds to Eqs.~(13) and (14) of \cite{Jakovac:2003ar}, just that we used the relation $\displaystyle \frac{\partial}{\partial n}\Big(\Li_n(-e^z)+\Li_n(-e^{-z}\Big)\Big|_{n=0} = -\gamma-\ln(2\pi)-\big[\Psi\big((1+iz/\pi)/2\big) + \Psi\big((1-iz/\pi)/2\big) \big]/2,$ which can be proven by comparing the high temperature expansion used there with the one given in \cite{Ayala:2014jla}.}:
\begin{subequations}
\bea
m^2_\textrm{eff}&=&m^2+\left(\frac{\lambda}{72}+\frac{g^2}{12}N_c\right)T^2+\frac{g^2 N_c}{36\pi^2}\mu_B^2, \label{Eq:m2_eff}\\
\lambda_\textrm{eff}&=&\frac{\lambda}{6}-\frac{g^4N_c}{8\pi^2}\left[\Psi\left(\frac{1}{2}+i\frac{\hat \mu_B}{6\pi}\right) + \Psi\left(\frac{1}{2}-i\frac{\hat\mu_B}{6\pi}\right) \right. \nonumber\\
  &&\left. + 2 + 2\ln\frac{4\pi T}{M_0}\right].
\label{Eq:lambda_eff}
\eea
\label{Eq:eff_coupl}
\end{subequations}
In the expressions above $\Psi(x)$ is the digamma function, $\hat\mu_B=\mu_B/T,$ $N_c$ is the number of colors, $g=m_q/\Phi_0$ (with $m_q=m_N/3$ and $\Phi_0=f_\pi/2$) is the Yukawa coupling between the pion and sigma mesons and the constituent quarks, and $m^2$ and $\lambda$ are the renormalized mass and the self-coupling in the $O(N)$ symmetric mesonic sector of the CQM model, which at the value $M_0=886$~\MeV~of the renormalization scale take the values $m^2=-326054$~$\textrm{MeV}^2$ and $\lambda=400.$

For $\mu_B\ge0$ the model exhibits a second order chiral phase transition line in the $\mu_B-T$ plane, which is obtained from the condition $m^2_\textrm{eff}=0$. This line of second order points ends in a tricritical point with coordinates determined by $m^2_\textrm{eff}=\lambda_\textrm{eff}=0.$ For $\mu_B>\mu_B^\textrm{TCP}$ the chiral phase transition is of first order and $m^2_\textrm{eff}=0$ gives the location of the first spinodal down to $T=0.$ The merit of the expressions in \eqref{Eq:m2_eff} and \eqref{Eq:lambda_eff} is that the line of second order phase transitions, which is actually an ellipse in the $\mu_B-T$ plane, can be determined analytically together with the location of the TCP. This makes the analytic continuation very simple, as we just have to change $\hat\mu_B^2\to-\hat\mu_B^2$ in the expression
\be
T_c(\hat\mu_B^2) = \sqrt{\frac{-72 m^2}{\lambda+6g^2N_c\big(1+\hat\mu_B^2/(3\pi^2)\big)}},
\label{Eq:Tc-model}
\ee
obtained from \eqref{Eq:m2_eff}, to go from real to imaginary chemical potentials.

\begin{figure}[!t]
\includegraphics[width=0.495\textwidth]{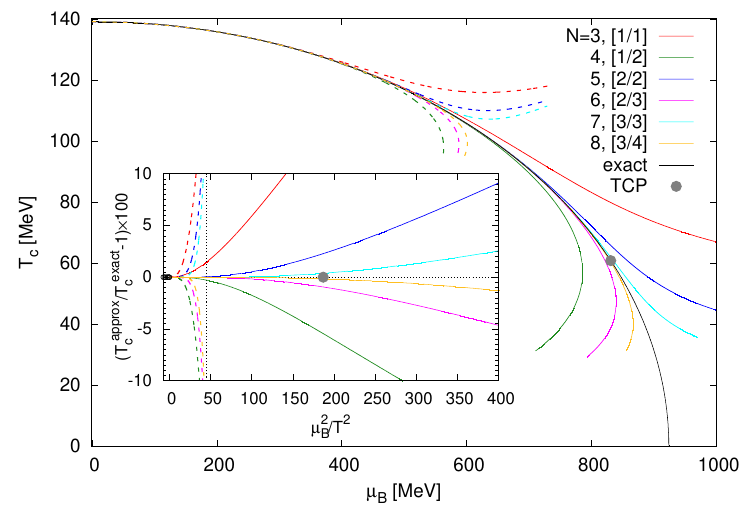}
\caption{Apparent convergence of the multipoint Pad{\'e} approximants $C_N$ (solid lines) determined from $T_c$ values at $\mu_B^I$ in the CQM model in comparison to the Taylor expansion of order $N-1$ around $\mu_B=0$ (dashed lines). In the main plot the parametric curves for the Pad{\'e} approximants are obtained as $(\sqrt{\hat\mu^2_B} C_N(\hat\mu^2_B),C_N(\hat\mu^2_B)).$ The bivaluedness of the subdiagonal approximants (and the curves of the odd-order Taylor expansion) reflects that $\mu_B:=\sqrt{\hat\mu^2_B} C_N(\hat\mu^2_B)$ has a maximum. The inset shows the percentage difference between the approximated and exact values of $T_c,$ with the vertical dotted line indicating the radius of converges of the Taylor expansion.  \label{Fig:model}}
\end{figure}

We can sample $T_c$ at imaginary values of $\hat\mu_B$ and fit multipoint Pad{\'e} approximants to the sampled data (see Appenix~\ref{app:mP}). Then, we can evaluate the Pad{\'e} approximant at real values of $\hat\mu_B$ and compare the value of analytic continued $T_c$ with the exact values obtained from \eqref{Eq:Tc-model}. This comparison is presented in Fig.~\ref{Fig:model}, where the inset shows the percentage difference between $C_N(\hat\mu_B^2)$ and $T_c(\hat\mu_B^2)$ using the same interpolating points as those shown in Fig.~\ref{Fig:prior} in the case of the QCD data. The main figure shows that the diagonal sequence converges from above, while the subdiagonal sequence converges from below to the line $m^2_\textrm{eff}=0$, which for $\mu_B<\mu_B^\textrm{TCB}$ is the line of critical points and for $\mu_B>\mu_B^\textrm{TCB}$ is the first spinodal. Given that the sampling range is $\hat\mu_B^2\in(-7.35,0],$ the accuracy of the $[3/4]$ Pad{\'e} approximant around the location of the TCP is remarkable, even though  $T_c(\hat\mu_B^2)$ is a rather simple function, as it represents an ellipse. This is even more so when one compares to the radius of convergence of the Taylor expansion around $\mu_B=0$ which is $\hat\mu_B^2\approx 44.2$, as given by the pole in \eqref{Eq:Tc-model}. We only refer to the TCP because $\mu_B$ (or $\hat\mu_B^2$) is rather large there; the Pad{\'e} approximant does not know about the existence of the TCP, as this is encoded in the quartic part of the tree-level potential and the second derivative of $\langle\bar q q\rangle/\Phi$ ($q$ is the constituent quark field) with respect to $\Phi$, which jointly determine $\lambda_\textrm{eff}.$

\subsubsection{Effect of the error}

Next, we investigate what happens when analytic continuation is performed in the presence of noise. As a reference point we start by generating ${T_c^i}$ configurations with a normal distribution characterized by mean calculated from \eqref{Eq:Tc-model} and standard deviation corresponding to the relative error $w_{T_c^i}/T_c^i=1\%$ and investigate to what extent should we decrease the relative error in order to get close to the curves obtained in Fig.~\ref{Fig:model} in the absence of noise. Note that $w_{T_c^i}$ is by a factor of two larger than the average error of the QCD data at imaginary chemical potential.

\begin{figure}[!t]
\includegraphics[width=0.485\textwidth]{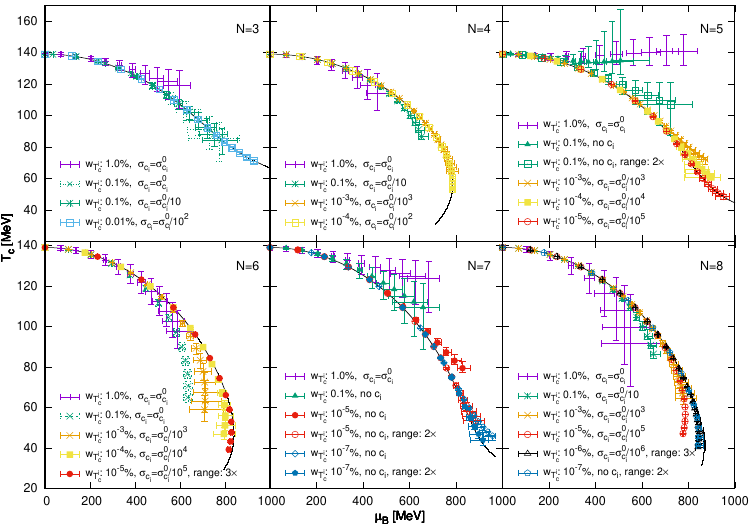}
\caption{Result of a mock analysis showing the effect of the error of the data and of the sampling range on the quality of the analytic continuation obtained via multipoint Pad{\'e} approximants of various order and by including or omitting information (the latter is denoted by 'no $c_i$' in the key) on the error of the Taylor coefficients in the evaluation of $\chi^2.$ For additional information see the main text. \label{Fig:modell_err}}
\end{figure}

We determine the coefficients of the multipoint Pad{\'e} approximant for each generated configuration, evaluate the approximant for positive $\hat\mu_B^2$ and, using the Bayesian method presented in Sec.~\ref{ss:bayes}, calculate $\chi^2$ including or omitting information on the Taylor coefficients, and then study the posterior distribution of these values. The method is applied to the QCD data in the next subsection, where it is referred to as Method 1. The control points used to calculate $\chi^2$ have $\hat\mu_B^2$ corresponding to the QCD data at imaginary chemical potential and $T_c$ obtained from \eqref{Eq:Tc-model}. We use a unique relative error of $T_c$ at all interpolating and control points, whose value is indicated in the key of Fig.~\ref{Fig:modell_err} ($w_{T_c^i}$ used to generate ${T_c^i}$ instances is twice the indicated value). When Taylor coefficients $c_1$ and $c_2$ are also used in the calculation of $\chi^2$, their values $c_1=-1.575$ and $c_2=0.0267$ is determined from the Taylor expansion of \eqref{Eq:Tc-model}, as for the reference value of their error, indicated in the key of Fig.~\ref{Fig:modell_err}, we use the error of the QCD data obtained from Ref.~ \cite{Bazavov:2018mes}, namely $\sigma_{c_1}^0=0.626$ and $\sigma_{c_2}^0=0.627.$ The sampling points in the range $\hat\mu_B^2\in(-7.35,0]$ are those used previously to obtain Fig.~\ref{Fig:model}. We also investigate the effect of changing the sampling range for fixed value of the error by increasing the lower bound of the interval by the factor indicated in the keys of Fig.~\ref{Fig:modell_err}. In the modified range the interpolating points are equidistant from each other.

Worth noticing in Fig.~\ref{Fig:modell_err} is that in the presence of noise the bands for $T_c(\mu_B)$ can deviate above some value of $\mu_B$ from the curves of Fig.~\ref{Fig:model} by more than the estimated statistical error. This reflects the ill-posedness of the analytic continuation problem. However, even with the largest error used, the Pad{\'e} sequence converges up to $\mu_B\approx 600$~MeV. For the mathematically curious, we also show the effect of increasing the range of the interpolation points. As expected, convergence is accelerated by the increase of the sampling range.  This is of course not directly relevant for QCD, as the Roberge-Weiss transition puts a limit on the available range for the interpolation points.

\subsection{Analytic continuation of QCD data \label{sec:Tc-lattice}}

We apply he method presented in Sec.~\ref{sec:method} to the continuation of the critical line of the QCD in the $T-\mu_B$ plane. Our main focus is the study of the convergence of Pad{\'e} series of the form $[p/p]$ and $[p/p+1]$ constructed:
\begin{enumerate}
\item
based only on interpolating points (multipoint Pad{\'e} approximants), or
\item
using interpolating points and the expansion $f(t)\approx c_0 + c_1 t +c_2 t^2$ around $t=0,$ as explained in Sec.~\ref{ss:model} (Pad{\'e}-type approximant).
\end{enumerate}
We use the continuum extrapolated values of $T_c$ recently determined on the lattice at $\mu_B=0$ and seven imaginary values of $\hat\mu_B=\mu_B/T$, namely
\be
\hat\mu_B(j)=i\frac{j \pi}{8}, \qquad j=0,2,3,4,5,6,6.5,7,
\label{Eq:j_vals}
\ee
given in Table II. of \cite{Borsanyi:2020fev} and the Taylor coefficients $\kappa_2$ and $\kappa_4,$ appearing in the parametrization
\be
T_c(\mu_B)=T_c(0)\big[1-\kappa_2(\mu_B/T_c(0))^2 - \kappa_4(\mu_B/T_c(0))^4 \big],
\label{Eq:kappas}
\ee
also extrapolated to the continuum limit in \cite{Bazavov:2018mes}.

With the notation of Sec.~\ref{sec:method}, the (assumed) function $f(t)$, which corresponds to $T_c(\hat\mu_B^2)$, is known at seven points $\tau_j=\hat\mu^2_B(j)<0$, corresponding to $j\ne0$ in the list given in \eqref{Eq:j_vals}, and we also know $c_0=T_c(\hat\mu_B(j=0))$, as well as $c_1$ and $c_2$ in terms of $\kappa_2$ and $\kappa_4$. The values of $\kappa_2,$ $\kappa_4$ and $T_c(0)$ reported in \cite{Bazavov:2018mes} give through the explicit relations given in App.~\ref{app:c-kappa} $c_1=-1.878$ and $c_2=0.0451$ with errors $\sigma_{c_1}=0.626$ and $\sigma_{c_2}=0.627.$

\subsubsection{Numerical implementation of the Bayesian approach \label{ss:Bayes-implement}}

\begin{figure}[!t]
\includegraphics[width=0.495\textwidth]{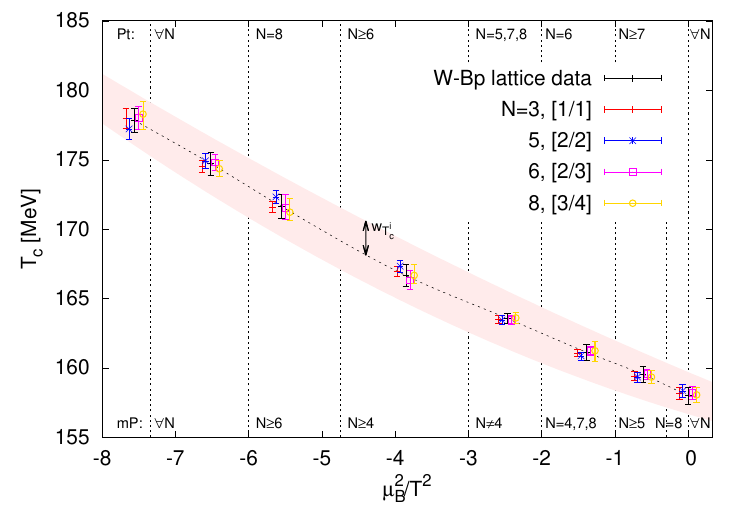}
\caption{Choice of the interpolating (node) points, whose position is indicated by the vertical dotted line, in comparison to the actual lattice data.  The labels indicate the use of the interpolating point in the multipoint Pad{\'e} approximant (bottom) and the Pad{\'e}-type approximant (top) of a given order, characterized by the number of independent parameters $N$. At the value of $\hat\mu_B^2$ corresponding to the lattice data points we show the mean and the error of the $T_c$ computed from the multipoint Pad{\'e} approximants generated with importance sampling (for the sake of the presentation the abscissa is shifted). The band indicates the standard deviation of the normal distribution used in \emph{Method 1} to generate $\{T_c^i\}$ instances, i.e. it indicates the prior distribution, excluding the factor that removes the spurious poles. The values $\bar T_c(\hat\mu^2_{B,\textrm{i}})$ at the node points $\hat\mu^2_{B,\textrm{i}}$ used in the expression \eqref{Eq:Pprior} of the prior are from the dotted curve in the band, which interpolates the mean values of the lattice data. }\label{Fig:prior}
\end{figure}

 In order to use the method presented in \ref{ss:model}, we need to generate $\{T_c^i\}$ instances at chosen interpolating points (also values of $c_1$ and $c_2$ in the case of the Pad{\'e}-type approximant) and then evaluate $\chi^2$, defined in \eqref{Eq:PP}, using the actual lattice data as control points. The interpolating points $\hat \mu_{B,i}^2$ used for the two types of Pad{\'e} approximants mentioned above are indicated in Fig.~\ref{Fig:prior}. The idea behind our choice was that each interpolating point of any of the used approximant fall in between two nearby lattice data points and be more or less equally distributed in the sampling range. The actual choice of the interpolating points is not important, however, in order to maximize the sampling range, one interpolating point is chosen close to the lattice data point with $\hat\mu_B^2(j=7)$ and, since we are interested in analytic continuation through $\mu_B=0,$ we also choose $\hat\mu_B^2(j=0)=0$ as an interpolating point.

We use two methods to generate input for the Pad{\'e} approximants. In the first method (\emph{Method 1}) we simply  generate $T_c^i$ from normal distribution with mean obtained by interpolating the mean of the lattice data and with the standard deviation (SD) indicated in Fig.~\ref{Fig:prior}. In this case $c_1$ and $c_2$, used in the Pad{\'e}-type approximant, are generated from a normal distribution with mean and SD given by Eqs.~\eqref{Eq:c_mean} and \eqref{Eq:c_SD}, respectively. As a result, $c_1$ and $c_2$ are taken into account in the calculation of $\chi^2$ only when using the multipoint Pad{\'e} approximant.
According to our prior, we only accept those configurations for which the corresponding Pad{\'e} approximant is free of spurious poles in the wide range $\hat\mu_B^2\in [-\pi^2,60\pi^2] $.
When using this method we calculate $T_c$ at some value of $\hat\mu_B^2$ as $T_c=R_m^n(\hat\mu_B^2)$ and the value of the real chemical potential as $\mu_B=\sqrt{\hat\mu_B^2}T_c$ and determine their percentile based error using the weight $e^{-\chi^2/2}.$

The second method (\emph{Method 2}) for generating input for the Pad{\'e} approximants is the importance sampling using the Metropolis algorithm with ``action'' $\chi^2/2$. The proposed value of $T_c^i$ in the Markov chain is generated using a normal distribution for the noise with vanishing mean and SD of ${\cal O}(1)$~MeV. In the case of the Pad{\'e}-type approximant we use normal distribution with standard deviation $\sigma_{c_i}, i=1,2$ to generate the noise for the Taylor coefficients $c_i$. Configurations for which the corresponding Pad{\'e} approximant has spurious poles in the range given above are excluded by assigning to them the value $\chi^2=\infty.$ For the remaining configurations $\chi^2$ is calculated using all the available lattice data according to the formulas in \eqref{Eq:PP}. The average and percentile based error of $T_c=R_m^n(\hat\mu_B^2)$ and $\mu_B$ for the Pad{\'e} approximants were calculated
in the standard way with the configurations provided by the Metropolis algorithm.

There are some peculiarities when doing importance sampling in this context. These are related to the spurious poles of the Pad{\'e} approximants, which appear as ``walls'' of infinite action in the Metropolis update. Configurations with spurious poles are not guaranteed to be isolated points in the space of all configurations, rather, there can be regions in configurations space where all approximants have a pole. One can easily stumble on an accepted configuration that is surrounded in most directions by configurations with a pole, thereby trapping the algorithm. To avoid this problem it is a good idea to mark out a temperature range sampled by the algorithm during the random walk and assign infinite value for the action if a proposed $T_c^i$ lies outside this range. I.e. even in the case of \emph{Method 2} a prior, like that in Fig.~\ref{Fig:prior} is used. An other reason to introduce this band is to exclude the Pad{\'e} approximants from having features in the interpolated range not present in the data, even if such an approximant has no pole and fits the data points acceptably. This is also a possibility, since Pad{\'e} approximants are rather flexible. In practice a two times wider band than the one shown in Fig.~\ref{Fig:prior} proved sufficient.

Another observation is that in some cases it was very hard to thermalize the system by updating the value of $T_c$ only at one interpolating point at a time. It proved more useful to propose in the Metropolis algorithm an updated array of $T_c$ values, as this procedure also substantially reduced the autocorrelation time.

\subsubsection{Results for the analytic continuation}

The first thing worth checking is the distribution of $T_c$ calculated from the Pad{\'e} approximants at $\hat\mu_B^2$ values corresponding to the actual lattice data points. For the majority of the approximants and lattice data points, the distribution is very close to a normal one with standard deviation compatible with that of the lattice data. The latter can be seen in Fig.~\ref{Fig:prior} in the case of the multipoint Pad{\'e} approximant, meaning that the selection of the $T_c$ instances based on $\chi^2$ works as expected. However, different low order approximants seem to select, withing the error, different ranges in the distribution of $T_c$ (and $c_i$ when the Pad{\'e}-type approximant is used). This is most visible in the case of the [2/2] approximant where the points posses a structure unseen in the lattice data. The multipoint Pad{\'e} approximant [1/1] is the most constrained by the likelihood, the error of $T_c$ being smaller than the lattice one, while the [3/4] approximant is the least constrained, matching closely the lattice error at all lattice points, and showing a wider range of the computed $c_1$ and $c_2$ coefficients. This loss of constraint is also reflected by the $\chi^2$ histogram whose pick moves to higher values when the number of parameters of the approximant increases.

\begin{figure}[!t]
\includegraphics[width=0.495\textwidth]{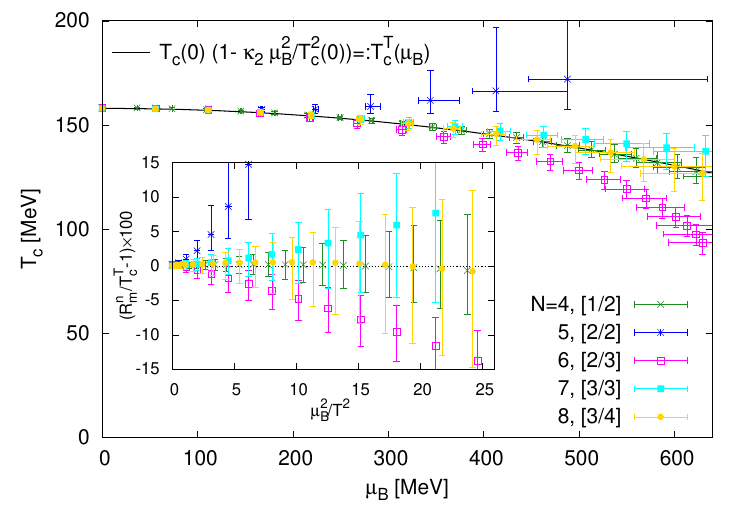}
\caption{Result of the analytic continuation via the diagonal and subdiagonal sequences of Pad{\'e}-type approximants constructed based on $\{T_c^i\}$ instances generated using importance sampling. The inset shows the percentage difference with respect to the $T_c(0)(1-\kappa_2 \mu_B^2/T_c^2(0))$ curve plotted in the main figure, for which we used $T_c(0)=158.01$~MeV and $\kappa_2=0.012$.\label{Fig:latTcont}}
\end{figure}

Now we turn our attention to the extrapolation. Since both \emph{Method 1} and \emph{Method 2} used to generate input for both the multipoint Pad{\'e} and Pad{\'e}-type approximants resulted in very similar results for the analytically continued $T_c(\mu_B)$ curve, we only present those obtained with \emph{Method 2} (importance sampling) in the case of the Pad{\'e}-type approximant constructed using the first and second derivative of $T_c(\mu_B)$ at $\mu_B=0.$ The fact that with \emph{Method 1} the analytic continuation does not depend on the approximant used, means that it makes no difference whether we take into account the Taylor coefficients only in the approximant or only in the calculation of $\chi^2.$ We remind that when importance sampling is used the Taylor coefficients are taken into account in the calculation of $\chi^2$ irrespective of the type of approximant, since otherwise the range in which $c_1$ and $c_2$ varies during the random walk would not be constrained.

Our main result on the analytic continuation is presented in Fig.~\ref{Fig:latTcont} in comparison with a simple parametrization of the crossover line based on the Taylor coefficient $\kappa_2.$ One sees that with the exception of the [2/2] type, the Pad{\'e} approximants tend to give smaller $T_c$ with increasing $\mu_B.$ Also, the behavior of the diagonal an subdiagonal sequences follow different patterns, similar to that observed in the model study in Fig.~\ref{Fig:model}. Apparently, the Pad{\'e} sequences converge, as, although [3/3] and [3/4] have overlapping error bars of similar size, the latter moves towards the band laid out by the [2/3] approximant. It remains to be seen  if this pattern survives the possible addition of new lattice data points, which will further constrain the fit, and/or an increase in the precision of the lattice data.

\section{Conclusions and outlook}

We presented a method for the numerical analytic continuation of data available at imaginary chemical potential that uses also the Taylor coefficients of an expansion around $\mu_B=0.$ Using lattice data that became available recently, we have investigated the continuation to real $\mu_B$ of the crossover line with a sequence of Pad{\'e} approximants, looking for apparent convergence as the number of independent coefficients increases. Such an analysis would have been less conclusive using the smaller data set available at imaginary $\mu_B$ in \cite{Bellwied:2015rza} and without taking into account the lattice data for the Taylor coefficients.

Our largest order Pad{\'e} approximants is very close to the simplest quadratic curve obtained with just the $\kappa_2$ coefficient. This means that if the observed apparent convergence is genuine, such a quadratic approximation might be applicable in a rather large range of $\mu_B$.
We would like to stress that, as discussed in the case of an effective model in Sec.~\ref{ss:model}, our results on the analytic continuation tell nothing on the possible existence and
location of the critical end point (CEP). It is also not possible to clearly determine the value of $\mu_B$ up to which the analytic continuation could be trusted.

The Taylor and imaginary chemical potential methods are usually considered to be competitors in the study of finite density QCD.
This is somewhat unfortunate, as the two methods tend to provide complimentary information. With the Taylor method, lower order
coefficients tend to be more precise, while data at imaginary $\mu_B$ tends to restrict higher order coefficients better, without
giving a very precise value for the lower orders. For the case of baryon number fluctuations, this can clearly be seen by comparing
Fig.~3 of Ref.~\cite{Borsanyi:2018grb}, where the signal for $\chi^B_6$ and $\chi^B_8$ is better, with Fig.~1 of Ref.~\cite{Bazavov:2020bjn},
where $\chi^B_4$ is much more precise.  This means that joint analysis of such data might be a good idea also for the equation of state, where
there are some indications---both from an explicit calculation on coarser lattices~\cite{Giordano:2019slo, Giordano:2019gev, Giordano:2020huj} and
phenomenological arguments~\cite{Connelly:2020pno, Mukherjee:2019eou, Connelly:2020gwa}---that the radius of convergence for temperatures close to the crossover is of the order $\mu_B/T \approx 2$, making a Taylor ansatz unusable beyond that point. This makes it mandatory to try different ansatze, or resummations of the Taylor expansion, and one possible choice could be the Pad{\'e} approximation method used here.

\begin{acknowledgments}
We would like to thank Sz.~Bors\'anyi, M.~Giordano, S.~Katz and Z.~R\'acz for illuminating discussions on the subject
and J.~G\"unther for providing the raw lattice data of \cite{Bellwied:2015rza} in an early stage of the project.
This work was partially supported by the Hungarian National Research, Development and Innovation Office---NKFIH grants
No. KKP126769 and No. PD\_16 121064, as well as by the DFG (Emmy Noether Program EN 1064/2-1). A.P. is supported by the J\'anos
Bolyai Research Scholarship of the Hungarian Academy of Sciences and by the \'UNKP-20-5 New National Excellence Program of
the Ministry of Innovation and Technology. In an early stage this research was also supported by the Munich Institute
for Astro- and Particle Physics (MIAPP) of the DFG cluster of
excellence ``Origin and Structure of the Universe''.
\end{acknowledgments}

\appendix

\section{The multipoint Pad{\'e} approximation method \label{app:mP}}
Following Refs.~\cite{Baker-Gammel} and \cite{VS_Pade}, we briefly summarize the construction of the multipoint Pad{\'e} approximant used to analytically continue functions known only at a finite number of points of the complex plane. In our case the continuation is done along the real axis, from negative to positive values.

When one knows the function at $N$ points $f_i=f(z_i),\,i=0\dots N-1$, the rational function approximating $f(z)$ is most conveniently given as a truncated continued fraction
\be
\label{Eq:Pade_CN}
C_N(z)=\frac{A_0}{1+}\,\frac{A_1(z-z_0)}{1+}\cdots\frac{A_{N-1}(z-z_{N-2})}{1}\,,
\ee
where we used the notation $\frac{1}{1+}x\equiv\frac{1}{1+x}$. The task is to determine the $N$ coefficients $A_i$ from the conditions $C_N(z_i)=f_i,\,i=0\dots N-1.$ Note that only $N-1$ values of $z_i$ appear in \eqref{Eq:Pade_CN}, $z_{N-1}$ appears in the condition $C_N(z_{N-1})=f_{N-1}$. The coefficients can be obtained efficiently as $A_i=g_i(z_i),\,i=0\dots N-1$, with the functions $g_i(z)$ defined by the recursion
\be
g_p(z)=\frac{g_{p-1}(z_{p-1})-g_{p-1}(z)}{(z-z_{p-1})g_{p-1}(z)}, \ \ 1\le p \le N-1,\ \
\label{Eq:recursion1}
\ee
with initial condition $g_0(z)=f(z),$ which means $g_0(z_i)=f_i$, when the function is known only in some discrete points. Working out explicitly the condition $A_i=g_i(z_i)$ for a few values of $i$, one sees that one needs to construct an upper triangular matrix $t_{i,j}$ using the recursion $t_{i,j}=(t_{i-1,i-1}/t_{i-1,j}-1)/(z_j-z_{i-1}),$ for $j=1,\dots,N-1$ and $i=1,\dots,j$, starting from its first row $t_{0,j}=f_j,$ $j=0,\dots,N-1$. The diagonal elements are the coefficients of $C_N$: $A_i=t_{i,i}.$ The relation of $C_N$ with the Pad{\'e} sequence is as follows: if $N\ge1$ is odd, then $C_N=[p/p]$ with $p=(N-1)/2$, while when $N\ge1$ is even, then $C_N=[p/p+1]$ with $p=-1+N/2.$

Writing $C_N(z)$ in the form $C_N(z)=N(z)/D(z),$ the numerator and denominator (at a given value of $z$) can be easily determined from the coefficients of the truncated continued fraction via the following three-term recurrence relation
\be
 X_{n+1}=X_n + (z-z_n) A_{n+1} X_{n-1},
\label{Eq:recursion2}
\ee
where for the numerator ($X=N$) one has $X_1=0,X_0=A_0$ and for the denominator ($X=D$) one has $X_1=X_0=1,$ and the iteration goes from $n=0$ up to and including $n=N-2.$ The coefficients $a_i$ ($b_i$) of the numerator (denominator) can be easily obtained by calling the above recursion \eqref{Eq:recursion2} at a finite number of points $z$ and solving a system of linear equations.

\section{Relating $c_{1,2}$ with $\kappa_{2,4}$ \label{app:c-kappa}}

In order to relate the coefficients $c_1$ and $c_2$ of the Taylor expansion $T_c(t)=c_0 + c_1 t + c_2 t^2,$ with $t=\hat\mu^2_B$, with the coefficients $\kappa_2$ and  $\kappa_4$ used by the HotQCD Collaboration, the expansion \eqref{Eq:kappas} has to be rewritten in terms of $\hat\mu^2_B:$
\bea
\frac{T_c(\mu_B)}{T_c(0)} = 1 - \kappa_2 \frac{T_c^2(\mu_B)}{T_c^2(0)}\hat\mu_B^2  -\kappa_4 \frac{T_c^4(\mu_B)}{T_c^4(0)}\hat\mu_B^4.
\eea
Then, using that $T_c^2(\mu_B)/T_c^2(0)\approx 1-2\kappa_2\hat\mu^2_B$ and $T_c^4(\mu_B)/T_c^4(0)=1+{\cal O}(\hat\mu^2_B)$,
we obtain
\be
c_1 = -\kappa_2 T_c(0) \qquad\textnormal{and}\qquad c_2= \big(\kappa_4-2\kappa_2^2\big)T_c(0).
\label{Eq:c_mean}
\ee
Ignoring the covariance between $\kappa_2$ and $\kappa_4$, which is not known to us, the error associated to these Taylor coefficients are
\begin{align}
\sigma_{c_1}&=\big[T_c^2(0) \sigma^2_{\kappa_2}+\kappa_2^2 \sigma^2_{T_c(0)}\big]^{\frac{1}{2}},\nonumber\\
\sigma_{c_2}&=\big[T_c^2(0)\big(\sigma^2_{\kappa_4} + 16 \kappa_2^2 \sigma^2_{\kappa_2}\big) + \big(\kappa_4^2+4\kappa_2^4\big)\sigma^2_{T_c(0)}\big]^{\frac{1}{2}}.
\label{Eq:c_SD}
\end{align}
In $c_1$ and $c_2$ and their errors we use the data of the HotQCD Collaboration also for $T_c(0).$

\end{document}